# Reviewing methods and assumptions for high-resolution large-scale onshore wind energy potential assessments

Russell McKenna[1], Stefan Pfenninger[3,4], Heidi Heinrichs[5], Johannes Schmidt[6], Iain Staffell[2], Katharina Gruber[6], Andrea N. Hahmann[9], Malte Jansen[2], Michael Klingler[6], Natascha Landwehr[6], Xiaoli Guo Larsén[9], Johan Lilliestam[8,10], Bryn Pickering[4], Martin Robinius[5], Tim Tröndle[8], Olga Turkovska[6], Sebastian Wehrle[6], Jann Michael Weinand[7], Jan Wohland[4]

[1] School of Engineering, University of Aberdeen, Aberdeen, United Kingdom
[2] Faculty of Natural Sciences, Centre for Environmental Policy, Imperial College London, London, United Kingdom
[3] Department of Engineering Systems and Services, TU Delft, Netherlands
[4] Institute for Environmental Decisions, Department of Environmental System Science, ETH Zurich, Zürich
[5] Institute of Energy and Climate Research – Techno-Economic Systems Analysis (IEK-3), Forschungszentrum Jülich, Jülich, Germany
[6] Institute for Sustainable Economic Development, University of Natural Resources and Life Sciences Vienna, Vienna, Austria
[7] Chair of Energy Economics, Institute for Industrial Production, Karlsruhe Institute of Technology, Karlsruhe, Germany
[8] Institute for Advanced Sustainability Studies (IASS), Potsdam, Germany
[9] Department of Wind Energy, Technical University of Denmark (DTU), Roskilde, Denmark
[10] Department of Economics and Social Sciences, University of Potsdam, Germany.

**Abstract**

The rapid uptake of renewable energy technologies in recent decades has increased the demand of energy researchers, policymakers and energy planners for reliable data on the spatial distribution of their costs and potentials. For onshore wind energy this has resulted in an active research field devoted to analysing these resources for regions, countries or globally. A particular thread of this research attempts to go beyond purely technical or spatial restrictions and determine the realistic, feasible or actual potential for wind energy. Motivated by these developments, this paper reviews methods and assumptions for analysing geographical, technical, economic and, finally, feasible onshore wind potentials. We address each of these potentials in turn, including aspects related to land eligibility criteria, energy meteorology, and technical developments relating to wind turbine characteristics such as power density, specific rotor power and spacing aspects. Economic aspects of potential assessments are central to future deployment and are discussed on a turbine and system level covering levelized costs depending on locations, and the system integration costs which are often overlooked in such analyses. Non-technical approaches include scenicness assessments of the landscape, expert and stakeholder workshops, willingness to pay / accept elicitations and socioeconomic cost-benefit studies. For each of these different potential estimations, the state of the art is critically discussed, with an attempt to derive best practice recommendations and highlight avenues for future research.

## 1. Introduction

Renewable energy has become an important player in global energy and development policy, representing 62% of new power generation capacity from 2009 to 2018 [1]. The most significant non-hydropower renewable technology is onshore wind energy, which has grown from 13% to 24% of the renewable energy capacity over this period [2]. To ensure climate protection and sustainable development, renewable energy capacities including wind must grow four times faster than this from now to 2030 [1].

Achieving such growth requires an accurate assessment of the potential available to achieve this objective in a cost-efficient and socially acceptable way. In this context, resource assessments for renewable energy have become an active field of research, driven by the worldwide push towards more sustainable energy systems. The significant attention devoted to this area in research over the past decades has led to substantial methodological improvements and more reliable resource estimates. This includes improvements in atmospheric modelling and data availability, land use mapping with open source data, as well as understanding of techno-economic turbine characteristics. One additional area which has seen particular methodological focus is improving the ways in which such studies account for non-technical (e.g., social) constraints for renewable resources like onshore wind (e.g. [7–10]).

Using a manual internet search and by screening 880 articles[1] and 88 reviews[2] in the Scopus database, we were able to identify and examine several previously published reviews on onshore wind energy. These include bibliometric analyses of general trends in this research area [11] or, for example, of specific factors that influence the economics of wind energy projects [12]. Other studies focus on the history of wind turbines [13] and global developments of wind energy diffusion in recent years [14]. A large stream of research deals with the forecasting of wind power generation or meteorological aspects, such as wind speeds, and has already resulted in many reviews [15–19]. Further reviews deal with onshore wind related to markets [20], environmental impacts [21], or detailed technical resource assessments of individual wind turbines in specific locations [22,23] such as urban environments [24]. Reviews of onshore wind potentials have mostly examined studies on specific aspects, such as the system integration of wind turbines, e.g., in electricity grid analyses [25] or energy system planning models [26]. There are also review studies that address onshore wind potential assessments in general, but usually only in a short section and mainly with a focus on the geographical potential [15,27,28]. Others have reviewed methods and tools for onshore wind potential assessments in the context of the broader spectrum of renewable resources, whilst focussing mainly on the technical aspects [29].

In summary, there is no review of best practices in identifying different (geographical, technical, economic) onshore wind potentials in large regions consisting of multiple countries or continents. At the lower end of the geographical scale, this review excludes detailed studies of wind park layout and planning (e.g. [30–32]) as such detailed analysis are not feasible at large scale. In addition, we similarly exclude sub-national or national resource assessments (e.g. [33–44]) that can be considered case studies and are therefore simply an application of standard state of the art methods reviewed here. The exception is when such studies employ a novel method, in which case they are considered with respect to these particular characteristics. Furthermore, the present review also considers non-technical, social aspects of onshore wind energy planning. Finally, the scope is limited to Horizontal Axis Wind Turbines (HAWTs), which are by far the most widespread due to higher aerodynamic efficiencies and lower costs than alternatives [45].

In the field of resource assessments for renewable energies, it is common to distinguish between different degrees of potential. Hence Hoogwijk et al. [46] distinguish four categories of potentials (cf. Table 1):

- The *theoretical or physical potential* refers to the total energy content of the wind within a specific region.
- The *geographical potential* equals the total area available for wind turbine installation accounting for technical, ecological and social constraints, such as minimum distances to infrastructure, protected areas or settlements.
- The *technical potential* corresponds to the wind power generated within the geographical potential. It considers constraints such as wind turbine characteristics, wind farm array losses and electrical conversion losses.
- The *economic potential* is the subset of the technical potential which satisfies criteria of economic profitability, which may differ between social welfare and private firm's profit-maximizing perspectives respectively. The economic potential strongly depends on prevailing energy-political and market frameworks.
- The above taxonomy can be extended further to consider that fraction of the technical potential considered practically achievable or desirable. So, for example, Jäger et al. [7] define the *feasible potential* as "the actual achievable economic potential, whilst accounting for market, organizational and social barriers, which mean that in practice the economic potential is not realized."

---

[1] Search query on 12/15/2020: *TITLE ("onshore wind" OR "wind power" OR "wind energy" AND (evaluat\* OR assess\* OR analy\* OR pot\* OR plan\* OR simul\* OR optimi\* OR model\*)) AND TITLE-ABS-KEY (wind AND (power OR generation OR energy) AND (evaluat\* OR assess\* OR analy\* OR pot\* OR plan\* OR simul\* OR optimi\* OR model\*) AND (potential OR locat\*) AND (generation OR cost OR lcoe OR econom\*)) AND SRCTYPE (j) AND (LIMIT-TO (DOCTYPE, "ar"))*

[2] Search query on 12/15/2020: *TITLE ("onshore wind" OR "wind power" OR "wind energy" AND (evaluat\* OR assess\* OR analy\* OR pot\* OR plan\* OR simul\* OR optimi\* OR model\*)) AND TITLE-ABS-KEY (wind AND (power OR generation OR energy) AND (evaluat\* OR assess\* OR analy\* OR pot\* OR plan\* OR simul\* OR optimi\* OR model\*) AND (potential OR locat\*) AND (generation OR cost OR lcoe OR econom\*)) AND SRCTYPE (j) AND (LIMIT-TO (DOCTYPE, "re"))*

Table 1 summarises how the above potentials are defined and relates them with examples to energy policy. Whilst theoretical/physical and geographical potentials are generally irrelevant for energy policy, technical, economic and feasible potentials are highly policy-relevant. On the one hand, these potentials are influenced by the laws, targets, and incentives at regional, national and international levels; on the other hand, these potentials and their dynamics arguably have a strong impact on policymaking, especially but not only in terms of the feasible potentials.

In this paper, we follow the same categorization of potentials, although we also highlight that this is an oversimplification, in particular the difference between geographical and feasible potential is vague. In the discussion in sections 5 and 6, we address these conceptual challenges in more detail. With the above framework of potentials as a structure, this paper provides an overview of recent developments in the field of resource assessment for onshore wind. In doing so, it discusses the state of the art in each of these areas and provides impetus for further research. Section 2 provides an overview of the geographical potential, especially the different land eligibility criteria employed in the literature on onshore wind energy. Section 3 then focuses on the technical aspects of onshore wind potentials, including the meteorological challenges and datasets, the technical characteristics of wind turbines, the issue of extreme wind speeds, and the spacing of wind turbines in wind parks. Subsequently, section 4 discusses economic aspects of onshore wind assessments, including the definition of the economic potential, the economic characteristics of turbines, various economic potential estimates for onshore wind, and the question of system/integration costs. Section 5 then turns to the so-called feasible potentials, reviewing the literature addressing those aspects not falling within a solely technical and/or economic framework, e.g., public acceptance, noise etc. Finally, section 6 critically assesses the methodological approaches presented in the preceding sections and presents some outlooks for further research.

*Table 1: Overview of different potential definitions and examples of their policy relevance.*

| Potential term | Definition | Policy relevance |
| --- | --- | --- |
| **Theoretical or physical** potential | Total energy content of wind, e.g. globally. | Generally irrelevant |
| **Geographical** potential | …the geographical area available for wind turbines, e.g. globally. | Generally irrelevant |
| **Technical** potential | Electricity that can be generated from wind turbines within the geographical potential, over a given period of time (e.g. a long-term average or an hourly time series over a specific year), and with a given turbine technology (e.g. current, future). | Wind industry R&D, innovation and market dynamics |
| **Economic** potential | Subset of the technical potential that can be realized economically. | Energy-political frameworks |
| **Feasible** potential | Subset of the economic potential after accounting for non-technical and non-economic constraints. | Public acceptance, market barriers, inertia / resistance |

## 2. Geographical onshore wind potential

The geographical potential of wind energy is mostly defined as specific geographical areas available to install wind turbines (e.g. [7,47–50]). Other names for this type of wind energy potential in literature are practical potential [51], preliminary area definition [52], environmental factor [53], generally suitable sites [10] or suitable construction area [54]. Some studies include the geographical potential in part or fully into the technical potential (e.g. [33,34,48]), while others do not even cover this step at all [35]. Other studies further divide the geographical potential, e.g. into geophysical suitability as well as technical and environmental permission [48]. In most studies, determining the geographical potential is one of the first steps in analysing the wind energy potential. However, this step can also be carried out after determining the technical potential (e.g. [51]).

This section provides an overview and justification of the constraints applied to determine the geographical potential and ranges of buffer/offset distances in literature are given (section 2.i). Subsequently, approaches to process the set of constraints and often used databases are listed with their main characteristics (2.ii).

### i. Criteria

The availability of specific areas for wind turbines is most often derived from a set of primarily geographical criteria. Set definition and the utilisation of criteria to determine geographic suitability differ in literature. In most cases these criteria are used as strict exclusion criteria with or without buffer distances (e.g. [7,48,50]) or combined into indicators like a suitability factor (e.g. [10,36,46]), which adds a quality criterion to the geographical potential beyond the mere binary exclusion of areas.

Table 2 gives an overview of typical criteria and the range of buffers applied to the geographical potential. Those criteria can be categorised into different types like physical or technical constraints (e.g., *slope*, *altitude* and *water bodies*), exclusion criteria in the context of the built environment (e.g., *settlements* or *roads*) and related legislation, and environmental constraints to protect flora and fauna. While some criteria like *settlements, protected areas, roads* and *railways* occur in several studies, others like *agricultural area* [49], *power plants* [48], *firing areas* [47], *glaciers* [37,55] and *tropical forests* [46] are applied only infrequently; the latter three result mainly from different characteristics of the studied regions. In addition, the criterium *forests* is handled differently in literature ranging from full exclusion (e.g. [38,39,49,51,54]) to allowing some shares being suitable for wind turbines (sometimes dependant on the respective scenario) (e.g. [36,46,51,55]). Furthermore, the listed criteria and their respective buffer distances are divided into sub-criteria in several studies depending on the availability of databases and their underlying level of detail and definition of categories like settlement types (e.g. [36,40,50]).

*Table 2: Overview of criteria applied to derive the geographical potential of onshore wind energy.*

| Criteria | Excludes | References |
| --- | --- | --- |
| Slope | Values above 1 – 30° | [10,36,37,40,48–50,55] |
| Altitude | Values above 2 – 3.5 km | [39,46,48–51,55] |
| Water bodies | Distances below 0 – 1 km | [7,37–39,42,48–50,54,55] |
| Settlements | Distances below 0 – 3 km | [7,9,10,34,36–40,42,46–48,50,53–55] |
| Roads | Distances below 60 – 500 m | [7,9,34,36–40,47–50,54] |
| Power Plants | Distances below 1 km | [48] |
| Airports | Distances below 1 – 6 km | [7,9,36,37,39,40,42,47–50,53], |
| Transmission lines | Distances below 60 – 250 m | [7,9,37,48,50,53], |
| Railways | Distances below 60 – 500 m | [7,9,36,37,39,40,47–50,53,54] |
| Protected areas | Distances below 0 – 2 km | [9,34,36–40,42,46–51,53,55,56] |
| Forests | Distances below 500 m | [38,39,49,54] |
| Tropical forests | Distances below 0 m | [46] |
| Glaciers | Distances below 0 m | [37,55] |
| Firing areas | Distances below 0 m | [47] |
| Sandy areas | Distances below 0 m | [50] |
| National borders | Distances below 3 – 50 km | [47] |
| Mining areas | Distances below 0 – 3 km | [7,50,54] |

Apart from studies explicitly focussing on urban areas, there is a general consensus about excluding settlements and in most cases employing offset distances [41]. Distances between dwellings and wind power installations are ensured mainly in two ways. First, immission control regulations enforce levels of noise and visual impacts to be below well-defined thresholds. Whether a turbine can be built in a certain location thus depends on the characteristics of the planned turbine and is decided on a case-by-case basis. Hence, some studies (e.g. [10,42,53]) list noise as dependent on distance from the wind turbine as a criterium in their geographical potential. Second, distances can be ensured by enforced setbacks, which in most cases are standardised but, in some cases, depend on the height of the wind turbine. Setback distances are usually much larger than necessary for immission control and therefore exclude larger areas from wind power installations. The amount of excluded area depends not only on the setback distance, but also on the definition of settlements. In Germany, for example, the available area for wind power installations at a setback distance of 1 km is reduced by more than 30% when setbacks are considered not only for pure settlement areas but also for areas of mixed-use [43]. Immission control thresholds vary between countries and setback distances are often defined on the subnational level; sometimes as low as the municipal level.

Wind resource assessments whose geographical scale is multi-national therefore have to include a plethora of different regulations, which is considered a challenging task. While most studies with sufficiently high geographical resolution consider setback distances to settlements, these setback distances are only rarely based on existing, actual regulation in the assessed regions of multi-national studies and instead are generic assumptions such as a uniform distance (e.g. 600 m) or a multiple of the tower height [8,9,44,57,58]. Whether the magnitude of setback distances has a large impact on study results likely depends on the settlement structure. For Germany, which has a high population density, the magnitude of setback distances can have a large impact on wind potentials, with the technical potential with a 1000 m setback being just 1/3 of the potential with a 600 m setback [59].

Even while several studies (e.g. [37,38,46,49,50,60]) include wind speeds in their set of criteria for the geographical potential, we classify this as belonging to the technical potential (cf. section 3). However, if the size of the covered area might result in computational challenges, excluding areas below a specific minimum wind speed can be a good way to overcome this computational barrier.

The arguments for selecting specific criteria and their buffer for the geographical potential range from technical, economic to societal and legal aspects. For example, the fall in wind power due to a reduction in air density is used to explain the exclusion of high altitude locations (e.g. [36,46,48,55]). Other examples are regional planning catalogues and existing legislations, which build the basis for buffer distances [9,36] or biodiversity and natural health, which are used as an argument to exclude protected areas [49]. A distinct argumentation is particularly important for criteria which either exclude large areas or whose overlap with other criteria is small. While the impact of criteria varies geographically, Ryberg et al. [61] show that forests, habitats, slopes, and settlements are most impactful and mining areas and airports are least impactful for studies in Europe. However, a quantification of the impacts on the reuslts probably cannot be derived for all criteria, and, therefore, some studies have started to incorporate surveys [62]. However, this issue seems to be more related to the feasible potential addressed in section 5.

Most often the set of criteria and their buffers are chosen once. Only some studies include further scenarios to explore the impact of different settings or future developments in the context of sensitvity analyis (e.g. [10,38,47,50,51]). Such scenarios typically add or remove restrictions and vary buffer zones to non-eligible areas or vary suitability factors. Hence, up to now most approaches for the geographical potential are more or less static.

### ii. Approaches and databases

Several studies (e.g. [38,48]) utilize only the previously selected criteria or combine them with an additional buffer distance to exclude further non-suitable areas. Hence, those studies interpret the criteria as distinction between eligible or non-eligible areas. In contrast, another type of studies applies suitability factors (e.g. [36,46,51,55]). Suitability factors are used for different purposes like to address uncertainty in the database due to a lack in level of detail [55] or to combine different level of details in databases [36]. These suitability factors typically range between 0 and 1 and are most often translated as the fraction of land eligible for wind turbines in a specific geographical category or grid cell (e.g. [46,50]). Besides suitability factors, applying fuzzy sets to define an acceptable level in terms of selected criteria, which are then combined into an integrated satisfaction degree via a multi-criteria decision making approach, is another approach in literature [42]. A combination of approaches, considering some cirteria as pure exclusion criteria and others via suitability factors or as fuzzy sets, exists in literature as well (e.g. [42,51]). Moreover, another study combines exclusion zones, economic viability and social acceptability into a suitability score [9]. However, we consider this type of score to belong more to the feasible potential types discussed in sections 5 and 6.

Even if the regional scope of studies on wind energy potential differ, some databases are frequently used due to their global or continental scope and their open availablity (Table 3). These databases are most often complemented with further national or regional databases including both open and closed data. These regional data can range from landuse data [7,10] to military air traffic lanes [63]. Furthermore, natural protected areas are also often defined by regional datasets [34]. The utilized databases can bear different spatial resolutions ranging from around 100m$^2$ to several km$^2$, whereby the lowest spatial resolution typically determines the level of detail of the wind energy potential analysis, with some studies combining several different databases (e.g. [7,47]). Nonetheless, only rarely a validation or uncertainty analysis is performed, implying that a dedicated analysis on the impact of using different types of databases is still missing.

*Table 3: Overview of global and continental databases utilized in determining the geographical potential.*

| Dataset | Classes | Openly Available | Spatial Resolution | Regional coverage |
|---|---|---|---|---|
| Corine land cover [64] | 44 | Yes | 100 m linear phenomena 25 ha areal phenomena | Continental |
| ESA Land Cover Climate Change Initiative [65] | 22 (compatible with GlobCover) | Yes | 300 m globally higher resolution Africa | Global includes yearly maps |
| Natura 2000 [66] | Sites designated under Birds Directive and Habitats Directive | Yes | Varying | Continental |
| EU's Common Database on Designated Areas [67] | Individually for each area | In most parts | Varying | Continental |
| World Database on Protected Areas [68] | Individually for each area | Yes | Varying | Global |
| Global 30 Arc Secon Elevation project [69] | Elevation | Yes | Lateral resolution ~1km at equator | Global |
| GlobCover land cover dataset [70] | 22 | Yes | 300 m | Global |
| Digital chart of the world [71] | Country border | Yes | - | Global |
| Geographical information system for the analysis of biodiversity data [72] | Biodiversity | Limited | - | Global |
| Moderate Resolution Imaging Spectroradiometer (MODIS) [73] | 5 different land cover classification schemes, primary land cover scheme with 17 classes defined by the IGBP | Yes | ≥ 500 m | Global |
| USGS HYDRO 1k elevation dataset [74] | Stream lines, basins, | Yes | 1 km | Global |
| Open Street Maps Project [75] | 28 primary features with various subfeatures | Yes | Varying | Global |
| NASA. SRTM 90m Digital Elevation Data [76] | Elevation | Yes | 90 m at the equator | Global |
| Google Earth [77] | Various | Varying | Varying | Global |

In terms of the accuracy of the employed geospatial databases, some studies use sources such as Open Street Map (OSM) to consider existing buildings. Whilst this open-source data is widely available, it differs greatly in its coverage. The OSM database is constructed with user-volunteered input, which naturally calls into question its completeness. For example, Barrington-Leigh et al. [78] assessed OSM's completeness of roads on average globally, concluding that roughly 80% of all roads are accurately represented in the database, a coverage which varies by country. In most European countries, the estimated road completeness is well above the global average, often around 99% complete, with the exceptions of Turkey (79%), Albania (75%), and, most notably, Russia (47%). In addition, Hecht et al. [79] estimated the completeness of buildings in several regions of Germany, and found significant discrepancies from known building locations. In the state of North Rhine-Westphalia buildings completeness was found to be 25%, while in the state of Saxony it was only 15%. For example, much more recently, Broveli and Zamboni [80] evaluated OSM building completeness in Lombardy Italy and found the dataset to be 57% complete. Another promising dataset in this context is the World Settlement Footprint, which has global coverage at 10 m resolution and to our knowledge has not yet been employed for global onshore wind potential analyses [81].

## 3. Technical onshore wind energy potential

This section discusses the technical potential of onshore wind generation, beginning with the meteorology (section 3.i) and wind turbine technical characteristics (3.ii), followed by a discussion of the influence of extreme wind events on wind power potential (3.iii) and wind turbine spacing in parks (3.iv). These aspects culminate in the technical potential, as defined in Table 1. Selected international studies are summarized in terms of technical and economic potentials for onshore wind in Table 6. The technical potentials in this table, based on about 20 cited studies, range from 96-580 PWh globally (up to 717 PWh including offshore) or 0.4-77 PWh for Europe. The latter is shown for selected studies in Figure 2 at the end of this section and the economic potentials are discussed in section 4.iii.

### i. Meteorological wind power potential

Broadly, there are two types of wind resource assessment that lead to two types of data requirements. First, there are static or climatological wind potential assessments, requiring a wind atlas with wind speeds and/or power densities. Second, there are models of time-resolved renewable generation variability for use in energy system modelling (e.g., [82–84]), requiring appropriate input data such as from wind masts or meteorological reanalyses. These two types of assessment can also be combined: for example, static products like the Global Wind Atlas can be used to bias-correct reanalysis-based time series [85–89]. Wind speeds increase with altitude through the lower atmosphere, which is typically modelled by a logarithmic or power-law relationship [90]. For example, capacity factors increase by 16-34% when moving from 50 to 100 metres above ground; and a further 8-15% when moving from 100 to 150 metres, averaged across several sites in Europe [91]. We now discuss five key sources of meteorological data in turn, which are summarized in Table 4.

**Observations.** Many wind speed observations are available from weather stations and masts, for example, via the UK Hadley Center HadISD database [92] and the Tall Tower Database [93]. Station measurements can be affected by relocation, device updates, measurement error, and changes in the local topography [94]. Using station measurements for large-scale studies of wind potential thus requires dedicated quality control procedures of the underlying data [92,93]. Measurements are also spatially and temporally irregularly sampled. For example, 51% of the 222 >10 m masts in the Tall Tower Database are in Iran, and none are found in South America or northern Africa [93]. There is scope to improve spatial coverage in future by including the growing number of deployed and long-running wind farms, assuming wind park operators are willing or forced to share their data. Due to their limited coverage and irregular sampling, observations can be of little use in large-scale wind power modelling efforts despite their undisputed value at specific locations or in statistical downscaling of modelled or reanalysis data.

**Global reanalyses.** Most large-scale studies and databases for wind power potentials rely on reanalyses (e.g., [87,95,96]). Reanalyses combine a numerical weather prediction model of the atmosphere with observations using a technique called data assimilation (e.g., [97]). They provide meteorological data on a global regular grid, with information considered representative for the entire grid cell. This differs from observational data which provides point-based information. The choice of a reanalysis-based product depends on modelling context, and which temporal and spatial scale needs representing. Well known reanalyses of the satellite era (1979 to today) are ERA5 [98] and MERRA2 [99]; ERA5 has also recently been extended back to around 1950 [100]. Several studies have been undertaken to assess the performance of reanalyses to capture wind speeds. Over flat terrain in Northern Germany and the Netherlands, global reanalysis results are relatively well correlated to measured data [101–104]. However, for MERRA-2 and ERA-Interim it is reported that high ramp rates are underestimated while lower ones are overestimated. Temporal variability in general is underrepresented in reanalysis [101], which is confirmed by Cannon et al. [105] particularly for individual locations. Ramon et al. [106] find important discrepancies with regard to interannual variability and decadal trends in satellite-era reanalysis, yet report that ERA5 agrees reasonably well with tall tower measurements, except in areas of complex terrain where the sub-grid orographic drag artificially lowers the simulated wind speeds [107,108]. In applications that require longer time series, centennial reanalyses like 20CRv3 [109] and CERA20C [110] are used to investigate long-term wind variability (e.g., [111]). However, there are documented deficiencies of these datasets, most notably strong wind speed trends in CERA20C that are likely spurious [112]. Another issue is that global reanalyses are relatively smooth and thus tend to exaggerate spatial correlations between neighbouring regions [85].

*Table 4: Overview of meteorological datasets' coverage and resolution. Coverage and resolution information is approximate and based on the given example datasets; other datasets exist which may sit outside the given ranges.*

| Type of data source | Example datasets | Coverage | | Resolution | |
|---|---|---|---|---|---|
| | | Spatial | Temporal | Spatial | Temporal |
| Observations | HadISD [92], Tall Tower Database [93] | Global (irregular) | Historical, 20-50 years (irregular) | Site-specific | 5min-1hr |
| Global reanalyses | MERRA-2 [99], ERA5 [98], JRA-55 [113] | Global | Historical, 40-70 years | 30-60km | 1-3 hrs |
| Long-term global reanalysis | 20CRv3 [109], CERA20C [110] | Global | Historical, 100-150 years | ca. 100km | 3 hrs |
| Regional reanalyses | COSMO-REA2 [114], COSMO-REA6 [115] | Regional | Historical, 7-22 years | 2-6km | 1hr |
| Wind-focused reanalysis | NEWA [107], DOWA [116] | Regional | Historical, 11-30 years | 2.5-3 km | 0.5-1 hr |
| Wind atlases | NEWA [116], GWA [117] | National to global | Historical average | 200-50m | N/A |
| Climate models | CMIP5 [118,119], CMIP6 [118], EUROCORDEX [119] | Global or regional | Historical and future, X-Y years | Ca. 10km - 300km | Hourly to monthly |

**Regional reanalyses.** While ERA5 provides hourly data with ~30 km horizontal grid spacing, higher resolutions may be required to resolve wind patterns in complex terrain [120,121]. In fact, using global reanalyses can lead to a severe underestimation of wind energy technical potential [122]. Regional reanalyses provide higher resolution. COSMO-REA2, for example, has a horizontal resolution of 2 km, and can effectively resolve meteorological phenomena from a scale of ~14 km [114]. This is sufficient to resolve some mountainous weather patterns [122], while disagreement with observations remains large in particularly complex terrain [103,122]. Downscaling is computationally expensive and sometimes leads to limited scope datasets; COSMO-REA2 only covers seven years and nine European countries. Since regional reanalyses are provided over a confined area, they rely on boundary data from a global reanalysis. As a consequence, potential large-scale issues in the global reanalysis can propagate to the regional reanalysis.

**Wind atlas datasets.** In contrast to current reanalyses that are not designed with a specific focus on wind energy, wind atlas projects like the New European Wind Atlas (NEWA) [107], Dutch Offshore Wind Atlas (DOWA) [116], the Wind Atlas for South Africa (WASA) [123] and the Global Wind Atlas (GWA) [117] provide tailored, long-term mean wind energy information at a high spatial resolution. NEWA is based on a dynamical downscaling of ERA5 using the WRF model evaluated against mast measurements and exists as a mesoscale and microscale product [107,124]. The spatial grid spacing of the mesoscale NEWA is 3 km at seven heights above ground level and provides wind speed and power density averaged over 1989 to 2018. The NEWA microscale atlas is based on a second linearized downscaling to 50 m spatial resolution [107]. The high-resolution details of the surface elevation and surface roughness are found to improve the long-term means when compared to observations [107]. However, higher resolution does not automatically mean higher quality [102].

**Climate models.** While reanalyses and observations are only available in hindsight, climate model projections can be used to investigate impacts of future climate change on wind power generation. Climate model simulations are fundamentally different from reanalyses and observations giving rise to different sources of uncertainty (e.g., [125]). Large ensembles of climate model simulations are available from the Climate Model Intercomparison Project (CMIP; [126,127]) and downscaled projections are available from the Coordinated Downscaling experiment (CORDEX) initiative [118,119]. These datasets have been used in different assessments related to future wind energy potentials (e.g., [128–136]). Pryor et al. [137] recently reviewed literature on the subject.

### ii. Wind turbine technical characteristics

The next stage of the analysis for the technical potential involves wind turbines, which are discussed in this section. The focus here is on horizontal axis wind turbines (HAWT) that adopt the lift principle, due to their higher conversion efficiency, greater reliability and economies of scale allowing for cost effective multi-MW machines [45,138].

HAWTs are not simply uniform and homogenous machines. Instead, the choice of generator and rotor are designed for the specific conditions they will experience, and so different turbine types will have very different performance characteristics. Wind speeds can be converted to power output using empirical power curves, statistical approaches, or physical meta-models. Empirical power curves are typically provided by turbine manufacturers [139], for example in [55,95,140,141], but these require appropriate smoothing to account for heterogeneity in wind speeds experienced at different turbines within a farm and at short timescales [142]. Statistical approaches take historical data for measured wind speed and power output, typically at a regional or national aggregation, to derive a relationship between the two which automatically accounts for smoothing and other factors (e.g. [105]). Hypothetical power curves can be derived using meta-models (e.g. [6,143]) based on turbine specifications such as the specific power. This can help with future-focused studies, as power curves for next-generation turbines typically only become available after they have been operational for some time.

Three key design factors which influence energy production are the turbine's capacity, its hub height (which influences the wind speeds experienced), and the ratio of generator capacity to blade length (which determines the specific power and thus the general shape of the power curve). Figure 1 shows the evolution of these three parameters over the past three decades for onshore wind turbines in Europe. Turbine capacity has increased 16-fold since 1990, with a steady increase of 106 kW per year on average. The dominance of the 2 MW platform is visible from 2005 through to 2013, but since then 3 and 3.5 MW turbines have become commonplace. Similarly, hub height now averages 100 metres, 2.5 times greater than in 1990. This has grown by 3 metres per year, but has plateaued since 2015.

In addition, the cut-in and cut-out wind speeds determine the feasible range of operation for a given turbine and thereby also the lower and upper bounds of wind speeds for actual power generation. Other technical characteristics affect the shape of a wind turbine's power curve and thus its productivity, including storm control (for safety), noise reduction settings (sometimes required in built-up areas), the assumed size of a wind farm, and technical degradation over the turbine's lifetime [144].

The specific power of a turbine is arguably the most important feature in determining a turbine's output. The blade length (rotor diameter) determines the swept area and thus how much wind energy the turbine is exposed to. The generator capacity determines the maximum rate at which energy can be converted into electricity. Over the past 30 years, the specific power of European onshore turbines has remained nearly constant, averaging $394 \pm 11$ W/m$^2$. The IEC categorises turbines by three wind speed classes, defined by the annual average wind speed they are suited for. For example, the Vestas V66/2000 (66m rotor diameter, 2000 kW generator) is a Class I turbine, suitable for sites with annual average wind speeds above 10 ms$^{-1}$. It has a specific power of 1.7 m$^2$/kW and would yield a capacity factor of 22.3% in central Scotland [91]. In comparison, the larger-bladed V80/2000 (Class II, 2.5 m$^2$/kW 398 Wm$^{-2}$) would yield 31.4%, and the V110/2000 (Class III, 4.8 m$^2$/kW 210 Wm$^{-2}$) would yield 47.9% in the same location. All are 2 MW turbines, but one produces twice as much energy as another. This simplified comparison overlooks the constraints on turbine spacing, however, as discussed in section 3.iv below.

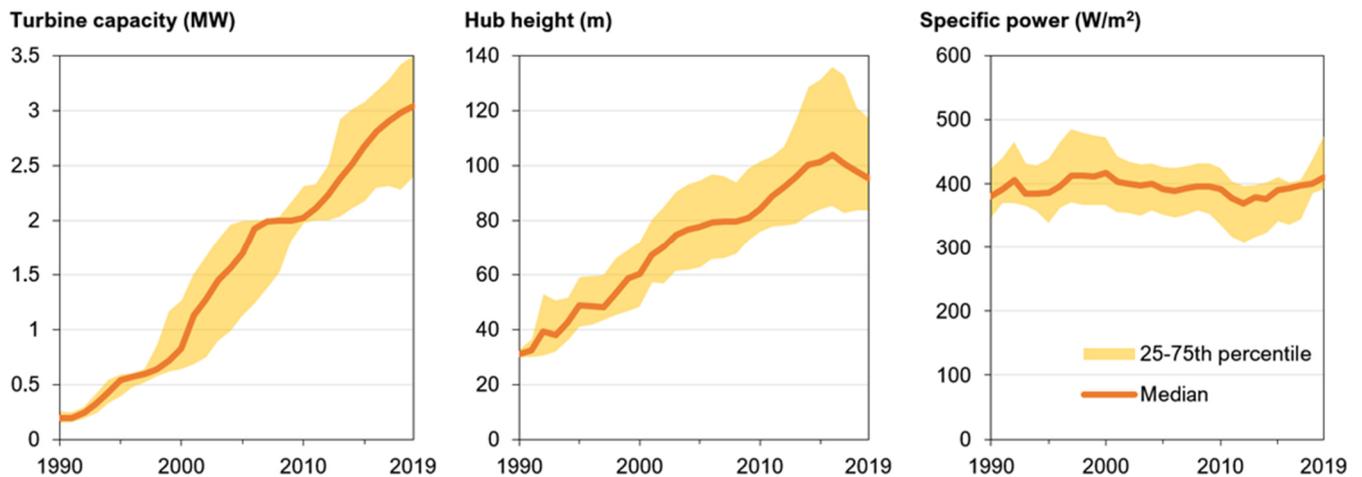

*Figure 1: Evolution of Europe's onshore wind turbines over the last three decades. Each panel shows the average specifications of new turbines installed in each year using data from Refs. [145,146]. Shaded areas represent half of all new turbines installed, covering the 25th to 75th percentiles.*

### iii. Extreme winds and their impacts on turbine design

As well as the general wind turbine characteristics discussed in the previous section, one specific and important characteristic is the ability to manage extreme wind conditions. For wind-farm planning, the expected extreme wind at hub height needs to be calculated to find suitable turbines that can harvest the most wind energy while also reducing the risk of damage from harsh wind conditions. For this reason, the fifty-year extreme wind at hub height is a design parameter specified in the IEC standard 61400-1 [139]. With climate change, and the resulting changes in frequency of extreme weather events, this may become an even more important issue than in the past.

The earliest dataset of fifty-year wind was produced in line with general civil-engineering applications [147]; each country used its own method, causing discontinuous values at national borders (e.g. [148]). Since then, there have been major methodological developments. Statistical algorithms have been derived to best represent the extreme wind samples from single or multiple types of extreme weather events (e.g. [149,150]). Pryor and Barthelmie [151] used the global ERA 5 data and calculated the 50-year wind at a height of 100 m, with a spatial resolution of about 30 km. Though a general issue is that the temporal and/or spatial resolution of numerical modelling data is often too coarse, Larsén et al. [152] developed the spectral correction method (SCM). This method uses wind variability that is generally missing in the modelled data over the relevant frequency range and uses information from limited available measurements or from a spectral model through the spectral domain. This method has been used to create an extreme wind atlas for South Africa [153], as well as for the whole globe [154].

Furthermore, tropical cyclone-affected areas have always been challenging for assessing design parameters. One example are the Chinese coasts: with measurements from 205 towers during 2003 and 2010, the ratios of the 50-year wind to the annual mean wind were calculated and can exceed the reference value by 5-10 times of the value given in the IEC61400-1 standard [16]. Using a hurricane conceptual model and best track data, Ott [155] calculated the 50-year wind for the west North Pacific. Larsén et al. [154] calibrated the SCM using the best track data and calculated the extreme wind for a tropical cyclone affected area in the northern hemisphere.

The estimation of extreme wind is still challenged by our understanding of flow across multiple scales, particularly in the range of a few kilometres to meters, the so-called spectral gap region (e.g. [156,157]). This limitation is reflected particularly in complex terrains and challenging severe conditions such as tropical cyclones and thunderstorms. It remains a problem to obtain reliable samples to assess the extreme wind climate and thereafter the distribution of these conditions when calculations cannot be achieved with high confidence. The implication for onshore wind resource assessments is that the technical potential is reduced, but as the extreme wind aspect is generally not considered, this represents a limitation in existing studies.

### iv. Wind turbine spacing and small areas

In addition to the technical wind turbine characteristics outlined in the previous sections, the wind turbine spacing in a wind farm strongly influences the technical generation potential. Industry practice for wind park siting and planning is to consider the extreme wind and other siting parameters such as turbulence intensity and load [139]. But this is generally overlooked in regional planning process and most of the studies reviewed here [5,158,159], due to a general lack of data. Private companies rely on commercial and confidential in-house calculation methods and data. The ongoing GASP (Global Atlas for Siting Parameters) project provides additional layers of publically available data to GWA at a spatial resolution of 250 m, including extreme wind, turbulence and wind turbine class over the globe [117,154].

Due to interference or wake effects, which reduce the output of wind turbines placed downwind of them, the capacity density of wind turbines is technically constrained. The explicit calculation of the optimal layout of wind parks is a combinatorial problem which is computationally hard to solve [160]. For this reason, the real world problem is often simplified to make it computationally tractable – for example, by discretizing the solution space of possible turbine locations [30] and applying evolutionary algorithms with heuristics [31] or particle swarm optimization [32] to optimize wind farm location and layout.

Another simple heuristic assumes a capacity density, as in e.g. [161]. Here, the number of wind turbines is simply constrained by an assumption on how much capacity, or how many turbines of a certain type, or how much rotor area can be placed on a given amount of land. A second option is to explicitly place turbines, using a rotor diameter distance heuristic [8,162]: by assuming that a certain distance between turbines has to be maintained, and that this distance depends on the size of the rotor, the number of turbines that can be potentially placed on a given stretch of land are determined by first placing a turbine and subsequently blocking land in the given minimum distance for further placements. Typical distances are in the range of 4D-7D [8], or 5D-10D [162], where D is the turbine rotor diameter and the larger offset is in the main wind direction, due to stronger wake effects. Nevertheless, some approaches assume the same distance factor in all directions (e.g. [8]), while others assume different distance factors, depending on the main wind direction (e.g. [6]). Wind roses which indicate the frequency of wind directions can be calculated from reanalysis products, but are also provided e.g. by the Global Wind Atlas [117].

When assessing the large-scale potential of wind energy, the impact of the upwind extraction of wind energy has to be taken into account for downwind wind parks due to wake effects. For example, Lundquist et al. [163] have found that these effects can have economically significant impacts up to 50 km downwind. When very large wind arrays are built, additional efficiency losses have to be expected, which can reach up to 70% for very large arrays (i.e. in the range of $10^5$ km$^2$) when turbine spacing is small [164]. In very large wind penetration scenarios which could be possible in the future, an overall reduction in wind speed should therefore be expected [137,164].

Relating to the employed turbine spacing, onshore wind resource assessments methodologies often must deal with many small areas (i.e. areas around 0.3-0.6 km² or smaller) resulting from the stepwise exclusion of unsuitable areas outlined in section 2. The application of turbine densities (in MW/km²) could be problematic [36] for these areas, as this could lead to a potential that is lower than that of a single turbine. In the literature there are basically two procedures for dealing with small areas. Firstly, the areas that are too small to build a single wind turbine could be aggregated and added to the total potential (as in [33,165]). This would overlook the shape of the areas as well as position with respect to one-another and therefore lead to an overestimation of the potential. On the other hand, the areas could be completely excluded. However, theoretically there could be enough area to build a single turbine, as only the area of the turbine tower base (around 0.1-0.2 km²) would be relevant for this. This approach would therefore lead to an underestimation of the potential [36]. Both of these effects can lead to an uncertainty of about +/-10% of the technical potential [36].

Overall, the approaches and assumptions employed to determine the technical potential for onshore wind outlined in this section lead to wide range in results. Figure 2 below shows the estimated potential for onshore wind in selected European countries, to our knowledge from the only studies with this broad scope and national disaggregation. The results between the four studies clearly show a large range, in some cases leading to national potentials that diverge by over 100%. Further discussion of differences in European-level results can be found in [166].

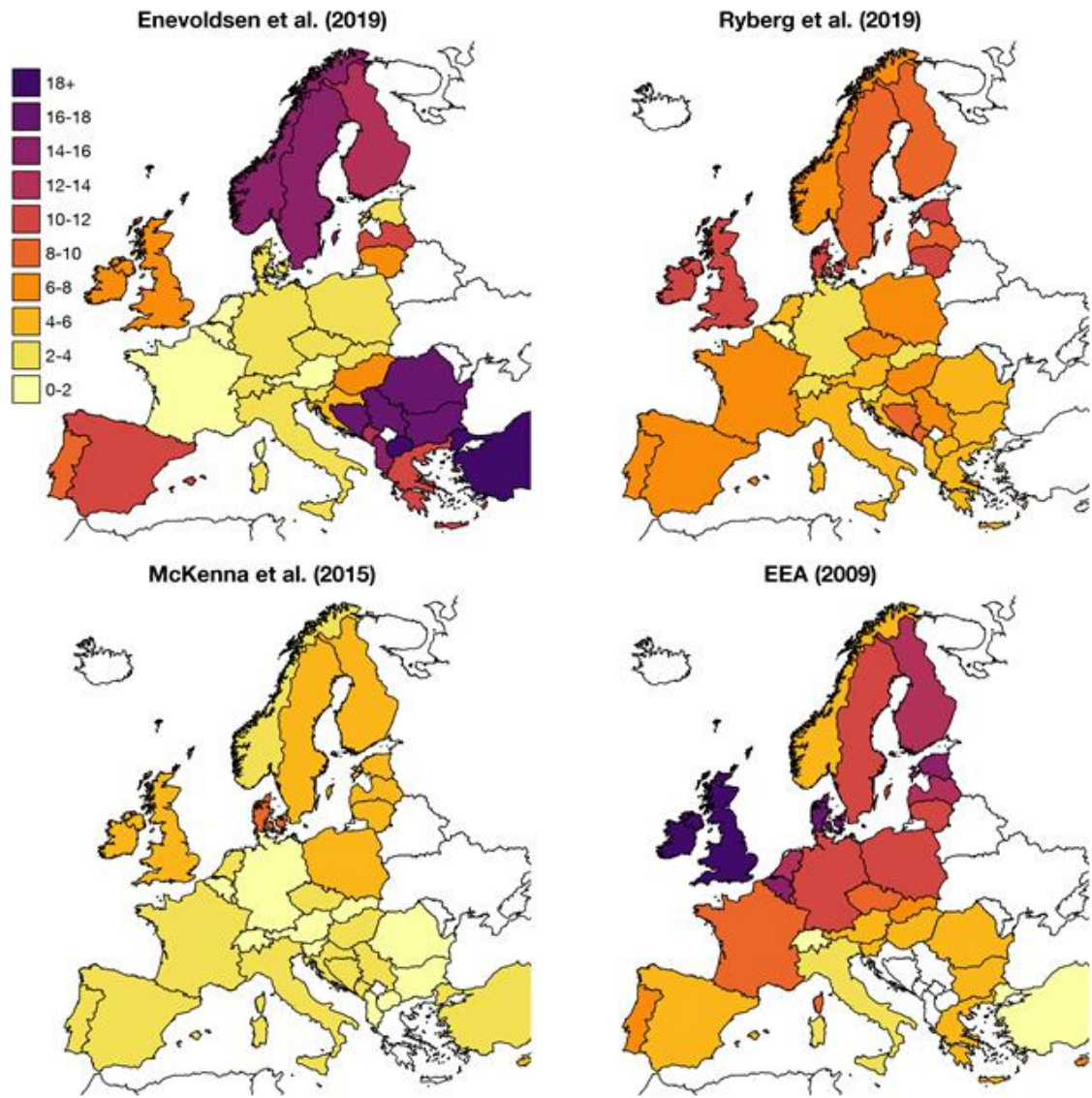

*Figure 2: Specific potential for onshore wind in selected European countries in GWh/km² of total land area [3–6]. To our knowledge, these are the only studies to have assessed this potential at the national scale across Europe. The shown potentials are technical, with the exception of Enevoldsen et al. [3], which refers to a so-called 'socio-technical' potential.*

### 4. Economic potentials of onshore wind

Having discussed the geographical and technical constraints relating to onshore wind resource assessments, this section addresses the economic dimension. It begins with a more elaborate definition of the economic potential and economic assessment criteria in section 4.i. Section 4.ii then presents economic characteristics of current and future HAWT technology, before section 4.iii discusses selected studies in terms of their (technical and in some cases) economic potentials. Finally, section 4.iv analyses the system integration costs of onshore wind and its implication for resource assessment methods.

### i. Defining economic potential and economic assessment criteria

In theory the economic potential is the fraction of the technical potential that currently can be economically realised. In practice, the term is not well defined, which limits studies' comparability and generalisability. A distinction between two different perspectives is relevant in this context:

- **The private economic or business perspective** assesses the economics of a wind turbine or park in the context of the prevailing market and energy-political framework conditions. This perspective assumes some prevailing market design (e.g. energy only spot markets) with perfectly competitive markets, and therefore does not consider any market-distorting effects of barriers or failures in the form of externalities, such as the noise impact or nearby settlements.
- **The public of welfare economic perspective** assesses the economics of increasing the share of wind power to some target, and hence looks at societal rather than the project-level economics. Externalities may also be considered in the analysis.

Both perspectives may employ similar economic metrics, e.g. Net Present Value (NPV), internal rate of return, or payback period. If the results are favourable (often in comparison to an alternative), the project or policy target is seen as economically attractive or 'profitable'. Although both perspectives assess the economic attractiveness of wind power, they differ in terms of the economic criteria employed. The different foci are for example reflected by the assumed discount rate for NPV calculations. Whereas the business perspective employs a private discount rate, often in the range of 8-15%, the welfare perspective uses societal discount rates, which are typically in the range 3-5% [167]. Because the discount rate very strongly affects the economics of capital-intensive technologies like wind power, the two perspectives generate very different results of the economic potential. Almost all studies reviewed here adopt the business perspective. We return to this perspective issue in section 5.

Of course, the business perspective – essentially the profitability of projects – depends not only on costs, but also on the revenue side, where the situation is changing from fixed-price to market premium schemes and auctions, and increasingly towards putting wind power on the general electricity market [168,169].

Assessing the economic potential is highly complicated, not only because economic parameters may vary widely across projects, but also because both costs and market conditions – and hence possible revenues – change over time. The impact of cost reductions on potential assesments can be profound. For example, recent studies have found that offshore wind at certain sites and in mature markets does not require subsidies [170], and because onshore wind generally has lower costs [171–173] the same will very likely apply to onshore wind too.

Consequently, many studies economically assess the technical potential, by employing some of the metrics outlined above in relation to discounted cash flow calculations. For example, a common approach is to relate the levelized costs of energy (LCOE) to the possible revenues, for example the average market price or achievable tariff [36]. These studies then find a potential generation that can be achieved at or below some particular cost. The LCOE comprises the total discounted costs over the the lifetime divided by the discounted energy production over the lifetime, and is a comprehensive metric to compare generation costs. But it ignores system integration and other external costs and because project-specific data is often lacking, it usually relies on generic assumptions for the cost of capital [174] (see next sections). Because of its intuitive simpicity, LCOE has become the dominant metric for costs of renewable power [173,175]. It often employs the weighted average cost of capital (WACC) concept [176], which accounts for the source and interest rates applied to the capital to finance the project.

### ii. Economic characteristics of turbines

Onshore wind energy is a near-zero marginal cost generator, which means that (almost) all of the cost is attributed to the construction of the asset, and 70-80% [177] of the costs are made up by the turbine itself. IRENA states the total installed costs (i.e. turbine, foundation, grid connection, etc.) in 2019 at 1,473 $\$_{2019}$/kW [173], whereas the U.S. Department of Energy sees turbine (only) costs of around 800 $\$_{2018}$/kW [178]. This is roughly comparable with the "2018 Cost of Wind Energy Review" by NREL, which assumes 1011 $\$_{2018}$/kW for the entire wind turbine, with the turbine rotor module accounting for 29%, the nacelle for 49%, and the tower for 22% of the cost respectively. The difference between the wind turbine and the total installed cost of 1470 $\$_{2019}$/kW arises from grid connection cost (32%), foundation (13%), construction and engineering (20%), engineering (8%) and financial cost (28%) [179].

Most sources however only provide the cost for the turbines and total costs and overlook some specific turbine characteristics, such as the drive train (i.e. gearbox or gearless), tower construction, tower height or rotor diameter. Table 5 gives an overview ot HAWT economic characteristics according to selected studies.

Markets for wind turbines strongly depend on regional and national energy-political and economic framework conditions, resulting in wind turbine prices being location-specific. Hence, modelling LCOEs can be carried out using a bottom-up approach, where all components such as blades, towers and balance of plant are costed, even going as far as indicating materials prices [180,181]. This is often carried out by academics and project developers, for example based on NREL's Wind Turbine Cost Model (LandBOSSE) [182]. This approach cannot reliably reveal the exact costs, however, due to the dynamic nature of turbine markets, their location-dependence and the reliance on privilieged, business-sensitive, and thus non-publicly available, information. This is addressed by analysts, such as the Bloomberg NEF wind turbine price index, which averages prices over many turbine types [183]. Hence, other approaches have evolved, which either cicrumvent the costing by using audited information to estimate costs [184] or using auction results [170,185], with both methods yet to be applied to onshore wind. Audits can be used to verify LCOE figures from the literature to match up "with the books" of publicly traded companies, based on the real costs of a projects from an accounting point of view. Auctions on the hand can provide an up-to-date proxy for LCOE in the near future, by estimating the underlying revenues of a particular project – and under the assumption of low cost margins, this is an estimate of the near-future LCOE.

The WACC has a large influence on the LCOE of renewables [186–188], which is potentially larger than the differences in CAPEX across countries. For onshore wind power in Europe, average WACCs vary strongly, from 3% in Germany to 11.7% in Greece, whereas the data for offshore wind is sparse, but range from around 6% in Germany and Belgium to over 12% in Great Britain [187]. Reducing the WACC from 7.5% to 5% reduces the LCOE by around 20%, so different assumptions will have a astrong effect on LCOE [173]. Hence, support schemes that lower project risk, such as feed-in tariffs have attracted low interest rate financing, enabling the relatively low WACC to contribute to the cost reductions in these technologies, alongside reductions in capex and opex and an increase in the capacity factor over time. On the other hand, if general interest rates increase again after the economic crisis, wind power WACCS and hence LCOEs may increase again.

*Table 5 Overview of HAWT economic characteristics from selected studies. The specific year of currencies is indicated, if known from the source.*

|  | Total capex per kW | Turbine only per kW | Operation and Maintenance (O&M) cost | WACC | Year (reported year, if given, otherwise source year) | Region |
|---|---|---|---|---|---|---|
| IRENA [173] | $_{2019}$1473 |  | 0.0060-0.0200 $_{2019}$/kWh | 7.5% (real, OECD countries and China) 10% (real, rest of the world) | 2019 | Global average |
| US DoE [178] | $_{2018}$1470 | $_{2018}$700-900 | 29 $_{2018}$/kW/a (2010-2017) |  | 2018 | United States |
| Gass et al. [159] | €1700 | €1400 | €0.0018 /kW/a | 7% | 2013 | Austria |
| European Environment Agency [5] | €1000 |  | 4% of capex per year[1] (40.0 €/kW/a) | 7.8% (private, presumed real) 4% (public) | 2005 | Europe |
| European Environment Agency [5] | €720 |  | 4% of capex per year[1] (28.8 €/kW/a) | 7.8% (private) 4% (public) | 2020 | Europe |

|  | Total capex per kW | Turbine only per kW | Operation and Maintenance (O&M) cost | WACC | Year (reported year, if given, otherwise source year) | Region |
|---|---|---|---|---|---|---|
| European Environment Agency [5] | €576 | | 4% of capex per year[1] (23.0 €/kW/a) | 7.8% (private) 4% (public) | 2030 | Europe |
| IRENA [180]. | $800-1350 | | | | 2030 | Global average |
| IRENA [180]. | $650-1000 | | | | 2050 | Global average |
| NREL [179] | $_{2018}1470 | $_{2018}1011 | 44 $_{2018}/kW 0.012 $_{2018}/kWh | 5.0% (real) | 2018 | United States |
| NREL[179] | $_{2018}1065 | | 34.3 $_{2018}/kW | 5.0% (real) | 2030 Low Innovation | United States |
| NREL[179] | $_{2018}929 | | 39.0 $_{2018}/kW | 5.0% (real) | 2030 Median Innovation | United States |
| NREL[179] | $_{2018}795 | | 43.6 $_{2018}/kW | 5.0% (real) | 2030 High Innovation | United States |
| NREL [189] | $_{2020}1436 | $_{2020}991 | 43 $_{2019}/kW/a 0.012 $_{2019}/kWh | 6.3% (nominal) 3.7% (real) | 2019 | United States |
| Danish Energy Agency and Energinet [190] | €_{2015}1330 | €_{2015}890 | 34.1 €_{2015}/kW/a 0.012 €_{2015}/kWh | | 2015 | Denmark |
| Danish Energy Agency and Energinet [190] | €_{2015}1120 | €_{2015}710 | 18.7 €_{2015}/kW/a 0.006 €_{2015}/kWh | | 2020 | Denmark |
| Danish Energy Agency and Energinet [190] | €_{2015}1040 | €_{2015}640 | 16.8 €_{2015}/kW/a 0.005 €_{2015}/kWh | | 2030 | Denmark |
| Danish Energy Agency and Energinet [190] | €_{2015}980 | €_{2015}590 | 15.5 €_{2015}/kW/a 0.004 €_{2015}/kWh | | 2040 | Denmark |
| Danish Energy Agency and Energinet [190] | €_{2015}960 | €_{2015}580 | 15.1 €_{2015}/kW/a 0.004 €_{2015}/kWh | | 2050 | Denmark |

[1] This assumes a 20 year lifetime and that the 4% are applicable per year, which is not stated in the source.

Recent estimates find that the LCOE of onshore wind power has decreased by 39% from 2010 to 2019 [173]. Whether and how fast technology costs will continue decreasing is debated and uncertain [5,180,191]. In the past, cost predictions have in some cases been accurate, such as the finding of the expert survey of Wiser et al. [192] in 2015: the surveyed experts expected a 10% cost reduction, which is roughly what can be observed today, and adds credibility to their estimated overall cost reduction to 2030 of 25%.

### iii. Economic assessments for onshore wind

Economic assessments of onshore wind have been a focal point of many European studies, as shown in Table 6 below. For example, the European Environment Agency [5] estimated the economically competitive potential of 9.6 PWh/a at a price of 55 €_{2005}/MWh or lower in 2020 and 27 PWh/a in 2030 and almost 60% of the total unrestricted potential. However, cost assumptions quickly become outdated (see previous section), which in turn likely increases the economic potential using current, lower costs.

Many studies define the economic potential as the capacity (or energy production) under certain cost thresholds (e,g, EEA [5], Ryberg [6]). Differences are identifiable in the approach taken and more crucially, cost estimates and projections for onshore wind influence the results significantly. The use of scenarios up to 2050 by some (e.g. IRENA [180]) is an alternative approach to evaluate the economic potential compared to the technical potenial.

In addition, forward-facing studies on a global scale, with time horizons of up to 2050 attempt to estimate future growth within economic and socio-economic contraints of each market, pointing towards large growth potentials in Africa, Asia and North America [180], which in some cases comment on socio-economic value [180,193].

Overall, then, the economic potential for onshore wind based on the studies in Table 6 ranges from about 7-110 PWh globally and around 5-27 PWh in Europe. Only a small number of studies in Table 6 include technical and economic potential [5,6,46,51,62], making generalizations from this small sample difficult. These studies differ widely in their methodologies, assumptions and dates of publication, meaning these results should be understood in the context of a specific study.

*Table 6: Overview of selected multi-country studies with technical and (in some cases) economic potentials[3].*

| Source | Focus | Available area [M.km²] | Assumed turbine size [MW] | Power density [MW/km²] | Technical potential [PWh/a] | Generation costs | Economic potential definition | Economic potential [PWh/a] |
|---|---|---|---|---|---|---|---|---|
| Hoogwijk et al. [46] | Worldwide onshore wind, several potentials | 11.00 | 1.00 | 4.00 | 96.00 | ≥ 0.05 US$/kWh | 1) ≤ 0.07 US$/kWh 2) ≤ 0.06 US$/kWh | 1) 14.00 2) 7.00 |
| Archer and Jacobson [194] | Worldwide analysis, onshore and offshore wind potential | n.a. | 1.50 | 6.00 | 627.00 | 0.03 - 0.04 US$/kWh | n.a. | n.a. |
| EEA [5] | European analysis, onshore and offshore wind for the EU27 | 5.40 | 2.00 | 10.00 | 41.00 | 1) 2020: 0.05 - 0.07 €/kWh 2) 2030: 0.04 - 0.06 €/kWh | ≤ 0.06 €/kWh 1) 2020 2) 2030 | 1) 8.90 2) 25.10 |
| Resch et al. [195] | European analysis, Several renewable technologies, potentials and costs for the EU27 | n.a. | 2.00 | n.a. | 0.40 | 0.05 - 0.10 €/kWh | n.a. | n.a. |
| Lu et al. [196] | Worldwide analysis, onshore and offshore wind potential | n.a. | 2.50 | 8.93 | 690 | n.a. | n.a. | n.a. |
| Held [197] | European analysis, Several renewable technologies for the EU27, considering social acceptance: here onshore wind in 2050 | n.a. | 2.00 | 3.00 | 1.96 | 0.05 - 0.13 €/kWh | n.a | n.a. |
| Scholz [198] | European and MENA countries analysis (40 regions), potential and | n.a. | 2.00 - 5.50 | 10.40 | 9.00 | 0.04 - 0.20 €/kWh | n.a. | n.a. |

---

[3] Includes studies with at least two countries, one whole continent or a global scope, based on the following search query on 14/02/2021: TITLE ( "onshore wind" OR "wind power" OR "wind energy" AND ( evaluat* OR assess* OR analy* OR pot* OR plan* OR simul* OR optimi* OR model*)) AND TITLE-ABS-KEY ( wind AND ( power OR generation OR energy ) AND ( evaluat* OR assess* OR analy* OR pot* OR plan* OR simul* OR optimi* OR model*) AND ( potential OR locat* ) AND ( generation OR cost OR lcoe OR econom* ) AND ( glob* OR euro* OR africa* OR america* OR australia* OR asia* OR world*)) AND SRCTYPE ( j ) AND ( LIMIT-TO ( DOCTYPE , "ar" ) )

| Source | Focus | Available area [M.km²] | Assumed turbine size [MW] | Power density [MW/km²] | Technical potential [PWh/a] | Generation costs | Economic potential definition | Economic potential [PWh/a] |
|---|---|---|---|---|---|---|---|---|
| | costs for renewable energy technologies | | | | | | | |
| Jacobson and Archer [199] | Worldwide analysis, onshore and offshore wind potentials | n.a. | 5.00 | 11.36 | 72 TW | n.a. | n.a. | n.a. |
| Zhou et al. [51] | Worldwide analysis, onshore wind potentials and costs | n.a. | 1.50 | 5.00 | 400 | n.a. | < 0.09 US$/kWh | 119.5 |
| Stetter [200] | Worldwide analysis, several renewable technologies | n.a. | 1.95 - 5.50 | 10.40 | 684 - 717 | 0.06 - 0.08 €/kWh | n.a. | n.a. |
| Mentis et al. [201] | African onshore wind, several potentials | 5.40 - 8.20 | 2.00 | 5.00 | 31.00 | n.a. | n.a. | n.a. |
| McKenna et al. [4] | European analysis, onshore wind potentials and costs | 0.74 | 3.00 | 8.3 - 18.6 | 20.00 | 0.06 - 0.5 €/kWh | n.a. | n.a. |
| Silva-Herran et al. [62] | Worldwide analysis, onshore wind potentials and costs | n.a. | 2.00 | 2.00 - 9.00 | n.a. | n.a. | 1) < 0.14 US$/kWh 2) < 0.10 US$/kWh | 1) 110 2) 29 |
| Bosch et al. [55] | Worldwide analysis, onshore wind potentials | 41.74 | 1.50 | 1.12 | 580.00 | n.a. | n.a. | n.a. |
| Eurek et al. [202] | Worldwide analysis, onshore and offshore wind potentials | 59.67 | 3.50 | 5.00 | 557.00 | n.a. | n.a. | n.a. |
| Dalla-Longa et al. [57] | European analysis, onshore wind potentials and costs | n.a. | n.a. | n.a. | 5.0-11.7 | n.a. | n.a. | n.a. |
| Enevoldsen et al. [3] | European analysis, onshore wind potentials | 2.71 | 4.50 | 10.70 | 76.52 | n.a. | n.a. | n.a. |
| Ryberg et al. [6] | European analysis, onshore wind potentials 2050 | 1.35 | 3.10 - 5.00 | 9.90 | 34.30 | 0.03 - 0.10 €/kWh | 1) ≤ 0.04 €/kWh 2) ≤ 0.06 €/kWh | 1) 4.62 2) 22.08 |

####    iv.    System integration costs

Most of the reviewed studies for potential assessment of onshore wind investigate only LCOE as the economic benchmark (e.g. [6,36]). These studies overlook important aspects of integrating non-dispatchable onshore wind into energy systems. According to Ueckerdt et al. [203] and Hirth et al. [204] the so-called *system LCOEs* include three additional cost components:

- *Profiling costs*, i.e. costs for additional dispatchable generation technologies to meet the residual load;
- *Balancing costs*, i.e. costs related to the deviation between forecast and actual non-dispatchable onshore wind generation;
- *Network costs*, i.e. costs for grid reinforcement and extension required to connect wind turbines to the network.

There are rudimentary estimates of the integration costs of onshore wind power for different contexts, valid for low penetrations. Typically, these studies analyze the short-run integration costs [203] by only considering the balancing and operational costs associated with marginal increases in installed wind capacities and generation. For this reason, they overlook more extensive long-run measures and their associated costs, such as network expansion and densification, which may either "happen anyway" or in the context of measures directed at wind energy integration; they also ignore wider developments in the power system. Therefore these rudimentary estimates of integration costs are only indicative and can only be applied to lower levels of wind energy penetration into the energy system. For example, a review by the European Wind Energy Association [5] revealed additional balancing and operational costs of 2.6-4.6 €ct/kWh for wind energy penetrations from 10-20% of gross demand. In addition, Heptonstall et al. have estimated that these costs are approximately 10-20 €$_{2017}$/MWh, for VRE penetration below 50% and with large uncertainties.

However, since these estimates depend on the specific system and are subject to significant uncertainty, the system LCOEs have to be explicitly considered in high-penetration assessments [204]. As wind (or PV) penetration increases, the stability of the power system is increasingly affected [205]. In Reichenberg et al. [206], for example, the system LCOEs increase sharply above 80% penetration (of wind and solar), especially due to the required strong expansion of transmission capacity. Whilst some studies take into account all the above-mentioned aspects of system LCOEs (e.g. [207]), the approaches to assess the technical potential of onshore wind do not. Nevertheless, initial approaches to include some of these effects have also already been developed for onshore wind potential assessments: in [208], wind turbines are clustered to wind farms and the grid connection costs (i.e. network costs above) are calculated for every technically feasible wind farm in UK showing that marginal and total costs more than double. The importance of system costs are further demonstrated by a recent study from China, showing that onshore wind power there has not reached grid parity[4] if system LCOEs are considered instead of LCOEs for generation only [212].

Nevertheless, there is still controversy about the costs of integrating wind power into electricity systems, including strong disagreement about how to measure them. Therefore, in future research, the results of sophisticated energy system models should be compared with economic approaches and empirical data from systems and markets where large wind penetrations already exist [213].

## 5. Feasible onshore wind potentials

The geographical potential (see section 2) is derived by excluding certain land areas for wind power development, for example, based on technical or legal constraints. The underlying assumption is that there is general agreement on these criteria, as implied for example by legislation. Although the geographic potential provides a good understanding about where *not* to build turbines, it does not imply that wind parks can be unquestionably deployed within the defined eligible lands. The required access to and use of land for wind parks may lead to conflicts with other land uses such as recreation, agriculture, subsistence, ecosystem services, or other renewable generation [214–216]. In addition, several externalities are associated with wind turbines, for example, related to noise, landscape and ecosystem impacts [217–219]. As a consequence, opposition against new projects in different world regions increased [220–222]. Therefore, land-use conflicts and externalities are increasingly considered in modelling wind power potentials [7,8,10,161,223–226]. In this paper, these studies are referred to as modelling the 'feasible' wind power potential (cf. Table 1).

---

[4] Defined as the equivalence of the LCOEs for a (decentralized) renewable generator with the purchase costs of electricity from the grid (including all grid fees and taxes). Once grid parity is achieved, the economic incentive to utilize own-generated electricity rather than import from the grid becomes stronger, in some cases justifying the investment in battery storage systems to increase this fraction [209–211].

Here, we first present methodological approaches which assess feasible potentials (section 5.i) and afterwards discuss which new data sources and indicators have been or might be used to represent these concerns in the modelling of wind power potentials (5.ii).

### i. Considering non-technical impacts in potential assessments

All wind power potential modelling approaches have decide which indicators are included, which thresholds are applied, and which buffer distances are used. Some parameters can be directly derived from legal or technical information. However, when modelling feasible potentials, this is mostly not the case. Therefore, recent modelling studies have relied on different approaches: the standard practice is that modellers choose the parameters (e.g. [7,208]) and/or stakeholders have been actively setting parameters in participatory approaches (e.g. [8]). Alternatively, patterns of wind power expansion have been transferred from regions with high levels of deployed wind power to regions with low levels (e.g. [161]), or acceptance and rejection of regulators has been used for statistical modelling of the likeliness of wind power projects being realized at specific locations [9,208]. Below we discuss how, once parameters are defined, feasible potentials are derived.

Mostly, *land-eligibility studies* (see Table 7-A) are used to assess feasible potentials. In contrast to traditional studies of geographical potentials, they consider a larger set of indicators related to conflicts in land-use and externalities. They apply a binary concept: land is assumed to be eligible or ineligible for the erection of wind turbines, depending on the chosen indicators. For instance, Jäger at al. [7] use assessments of landscape quality by citizens to differentiate between eligible and ineligible areas, resulting in a economic potential of 11.8-29.1 TWh in Baden-Württemberg, which is less than 50% of the technical potential, while Turkovska et al. [223] exclude all land-use with low anthropogenic activity from wind power development. This is a pragmatic approach, but it does not reflect the complexities of real siting decisions and particularly the trade-offs between different siting scenarios. In the following, we describe two alternative methodologies which transcend land eligibility studies.

The assessment of the *socially optimal expansion of wind power* seeks to identify one optimal spatial allocation of wind turbines. Therefore, the full social costs, arising, for example, from wind turbines' impact on the valuation of landscapes [227] or the environment [224], and benefits, for instance from lower integration cost of wind turbines compared to alternative sources of renewable electricity [228], need to be taken into consideration. Based on a quantitative valuation of social costs and benefits, allocations in terms of capacities and locations can be identified, which lead to the highest welfare gains or lowest welfare losses. To the best of our knowledge, a full welfare analysis has not yet been conducted, even though the social cost side has been explored (see Table 7-B). As the analysis of social welfare is concerned with the effects of wind power expansion on society as a whole, it is not suited to study individuals' or interest groups' patterns of technology acceptance. Moreover, welfare analysis is criticized for requiring comparisons of individuals' "happiness", which is possible only under the assumption of cardinal (i.e. meaningfully measurable) utility [229]. Furthermore, it is questionable whether a complete set of relevant welfare effects can be incorporated into actual, applied analysis. Conditional on these assumptions, welfare analysis allows to draw conclusions, for example, on potential compensation schemes for the ones affected by possible negative impacts of a specific wind turbine allocation.

In contrast, *multi-criteria analysis* (see Table 7-C) makes the trade-off between different objectives explicit. Eichhorn et al. [10], for instance, show the trade-off between minimizing wind turbine impacts on birds, maximizing their distance to settlements, and their total energy generation for different allocations of wind parks. As another example, Harper et al. [9] analyze the trade-off between social acceptance and the costs of wind power projects. They measure social acceptance by means of a statistical analysis that assesses which variables are correlated with regulatory acceptance or rejection of projects. Multi-criteria analysis can also be used in a multi-objective optimization framework. Instead of optimizing only for one objective, such as cost, several of them are optimized separately. This approach allows to explicitly derive so-called efficient frontiers, which indicate Pareto-efficient loci in the solution space and corresponding trade-offs [232]. For example, Drechsler et al. [233] analyze trade-offs between the equity of spatial distributions of wind turbines and the corresponding total system cost. Multi-criteria decision analysis aims at aggregating the different objectives to one joint indicator which can be used to directly inform policy making. The Analytical Hierarchy Process (AHP) [234] is frequently used for that purpose in wind power potential modelling studies. Here, experts or decision makers define relative weights for the different objectives in a structured process [60,231,235].

*Table 7: Examples of modeling approaches which determine 'feasible' potentials for onshore wind.*

| Modeling approach | Details | Region | Reference |
|---|---|---|---|
| *A - Land-eligibility* | | | |
| Participatory modeling | Exclusion zones and buffering zones are defined by input from stakeholder groups | Austria | [8] |
| Empirically observed saturation of wind power deployment | Characteristics of Austrian and Danish expansion are taken as basis for expansion in Czech Republic | Czech Republic | [161] |
| Landscape quality indicators | Regions with specific aesthetic landscape value, as measured by surveys and extrapolated to the whole state, are excluded | Baden-Württemberg, Germany | [7] |
| | Public landscape scenicness evaluation of crowd-sourced geotagged photographs for all of Great Britain | Great Britain | [208] |
| Avoidance of biodiversity impacts | Exclusion of all natural vegetation areas based on land-use and land cover maps | Brazil, Canada, global | [223,225,226] |
| Influence of wind turbines on property value | Acceptance costs derived from compensation/property purchase costs, property value loss and surveys | Denmark | [230] |
| *B Welfare analysis* | | | |
| Minimum social cost of wind power expansion | Total social cost of wind power development is minimized | West-Saxony, Germany | [224] |
| *C Multi-criteria analysis* | | | |
| Multi-criteria framework | Trade-off between three different objectives (bird collisions, settlement distance, energy performance) made explicit and aggregated to indicator. | Germany | [10] |
| Multi-criteria decision analysis | Trade-off between economic potential and social acceptance, as measured by statistically modelling the influence of variables on rejection rates of wind power projects | UK | [9] |
| Multi-criteria decision analysis | Analytical hierarchy process to derive weights for economic, social, environmental and technical criteria | Oman | [231] |

The data models, visualizations and maps employed in wind potential assessments make implicit arguments. As cognitive and normative devices, they constitute specific acts in the social production of space and are therefore deeply interwoven with power and knowledge [236,237]. As McCarthy and Thatcher [238] highlight in the context of the World Bank's renewable energy resource mapping initiative, a critical examination of spatial databases, key visual technologies and representations is important, by evaluating the geographical potential of rendered eligible land to attract global investment in wind power or other renewable energy production over other (perhaps traditional) land use systems [239]. Here, the hierarchization of indicators in data models also tends to underestimate the issue of land tenure insecurities, collective and informal property regimes, and the relevance of access to and control of 'marginal' lands for livelihoods, which is a major problem given the current spatial expansion of wind power in the Global South.

ii.     Employed data sources and indicators for feasible potentials

An understanding of the broader concerns surrounding wind power installations requires additional indicators for measuring related impacts. We discuss three interesting research avenues here, which have seen substantial contributions in recent years. They represent empirical research on the impacts of wind power installations and have not necessarily been applied to feasible potential studies yet (see Table 8). The first category of impacts relates to landscape quality, measured by direct indicators such as surveys on the willingness-to-pay or willingness-to-accept [224,230], by life satisfaction surveys [240], by surveys rating landscape quality [241], or by experiments measuring the physiological and behavioural reaction of participants to audio-visual impacts of renewable installations [242]. Changes in property prices [227,230] and decisions by regulatory authorities [243] have been used as proxies for the revealed individual or community preferences over wind power installations.

*Table 8: Example indicators for modeling social, environmental, and economic impacts of wind power generation.*

| Type of Impact | Data sources | Indicator | Reference |
|---|---|---|---|
| **Landscape quality** | Housing prices | Change in property prices | [227,230] |
| | Choice experiments | Willingness-to-pay, Willingness-to-accept | [224,230] |
| | Life satisfaction surveys | Life satisfaction (11-point Likert scale) | [240] |
| | Licensing decisions by regulators | Probability of licensing | [208,243] |
| | Photo rating experiments | Rating of landscape quality | [208,241] |
| | Measurement of physiological and behavioural reactions to renewable energy installations | Electrodermal activity | [242] |
| **Environmental impacts** | Remote sensing & GIS analysis | Direct land footprint and replaced land-use by the wind park | [223,264,265] |
| | GPS Tracking | Spatial movement corridors | [251] |
| | Maps on biodiversity | Natural habitat maps | [252–254] |
| **Land access** | Public GIS data | Property right information | [260] |
| | GIS data and remote sensing | (Traditional) land-use | [261,262] (not wind power related) |
| | Participatory maps | (Counter)maps claiming access to and use of 'undesignated' public lands by e.g. traditional and indigenous communities | [221] (not wind power related) |
| | Global environmental justice atlas | Conflicts over wind power projecs | [263] |

Another stream of research has assessed the environmental impacts of renewables. Traditionally, potential studies exclude environmentally protected areas and, sometimes, forests when defining the geographical potential. This practice, however, does not acknowledge the environmental vulnerability of land outside of these two categories. Savannas, shrublands, natural grasslands, amongst others, also provide a series of ecosystem services, such as stabilizing the local climate [244] and providing a natural habitat for endangered terrestrial species [245,246]. Multiple studies on various physical impacts of wind power showed that its development often occurs in such areas [223,247–250]. Furthermore, species movements, particularly bird movements, have been tracked with GPS devices to understand how they are affected by turbine placements [251], while natural habitat maps have also been used to assess the impact of wind parks on biodiversity [252–254].

Studies on opposition against wind power project in the Global South have shown that territorial conflicts and livelihood impacts resulting from land tenure insecurity, displacement processes, distributive injustices and missing options for financial and procedural participation emerge frequently [220–222,255–257]. In comparison, studies from the Global North emphasize more strongly the interrelations between socioeconomic (e.g. housing prices), environmental (e.g. biodiversity threat) and cultural (e.g. aesthetic cultural landscape values) effects of wind power [227,241,258,259]. We therefore consider the inclusion of ownership information [260], (traditional) land-uses [261,262], participatory mappings [221], and land conflict databases related to renewable infrastructures [263] into wind power potential studies, a very important research avenue.

## 6. Summary and conclusions
### i. Summary of existing methods

In this section we synthesise the main limitations of existing methods, where appropriate referring to best practice, based on the structure of sections 2-5 in the paper. In general, the analysis demonstrates a lack of consistency in the definition and application of the potential terms. Often the terms are employed incorrectly and in some cases are not employed whatsoever, so that the actual potential being analysed is not clear and may be misunderstood [3,166,266].

In terms of the **geographical potentials**, the main challenge lies in formulating a stringent definition of this potential. Once a set of criteria and buffer distances have been defined, they are generally applied to the whole region under consideration without any regional differentiation. This is problematic because it overlooks important regional differences in legal and political requirements that may strongly impact the onshore wind potential. On the other hand, potential estimates at the international scale are required to include a plethora of different regulations, which is a very resource-intensive and challenging task. For this reason, most resource assessments for onshore wind tend to be more or less static, reflecting the situation at a snapshot in time and overlooking temporal dynamics in framework conditions, technologies and costs – the exception here is obviously explorative studies that explicitly explore possible future developments in technology efficiency and costs (cf. Table 5). One further limitation relating to the geographical dimension of the analysis, is that a sensitivity analysis is only rarely performed. This means that the impacts of employing different databases, from a variety of time periods and at different spatial resolutions, on the results is not well understood. Exceptions here include Ryberg et al. [6], who did assess this and can be considered best practice in this regard.

In relation to the **technical potentials**, several aspects should be highlighted. There is a growing number of climatological datasets, some of them specifically targeted towards wind energy, which can be used to quantify mean wind power generation and its spatio-temporal variability. While these datasets generally provide a solid base to estimate wind energy technical potentials, no single dataset serves all purposes. Whether observations, reanalyses or a wind atlas provide the most suitable information often critically depends on the specific context. Cross-validation of results obtained using multiple climate data sources generally increases robustness of results given that biases and disagreements between different datasets exist. In terms of the technical turbine characteristics, some studies are backward-looking, using smaller and shorter turbines than are currently employed (e.g. [5] [159]); rather than forward-looking and considering the next-generation turbines that could be expected for the near-term future. Other problematic assumptions observed in previous studies include: assuming a single turbine type in all locations (e.g. [3] [159] [5] [202]), when in reality models with lower specific power are used in lower-wind locations; assuming a skewed power density (e.g. all Class III) which gives high capacity factor and thus energy yield (e.g. [3] [267]); and assuming single capacity factors for all locations (e.g. [3]), when there is notable heterogeneity across Europe. Assumed capacity factors range widely across previous studies, from 20-30% (1720-2630 full-load hours). In addition, extreme winds and turbulence intensities are typically not considered in studies of onshore wind potentials, despite being important siting parameters. This implies, at best, an inappropriate selection of turbine/IEC class for a specific location and, at worst, an overestimation of the energy yield. Turbine characteristics are often overlooked, with very simplified assumptions employing one or only a few different turbines. Turbine spacing is typically very rough and only a few studies actually place turbines in the landscape.

Next to the feasible potential, **the economic potential** is probably the most roughly defined. As well as the business and welfare economics perspectives, researchers have employed market energy prices and subsidy levels, as well as other thresholds to define their economic potential. The application of the LCOE as the main economic yardstick to compare and assess renewable technologies is problematic due to its limited scope, overlooking integration costs and other externalities. Therefore, in future research, the results of sophisticated energy system models should be compared with economic approaches and empirical data from systems and markets where large wind penetrations already exist [213]. These approaches also enable a departure from the purely business perspective that many studies adopt, in order to derive resource distributions for onshore wind energy that optimal from a welfare economics perspective.

The discussion in section 5 demonstrated the emergence of a relatively new research field over the past decade, which tries to go beyond purely spatial and techno-economic potential assessments and assess actually **feasible potentials** which might be realistically achieved in practice. This research is diverse in terms of the perspectives, methods and indicators adopted, but has in common the aim to apply quantitative methods to questions surrounding possible future locations of onshore wind energy generators. In economic terms, much of this research attempts to internalize externalities relating to wind power development, such as those concerning visual landscape impact, noise, and land use competition aspects. This new research thread also has in common the attention to the interaction between the technical system of wind energy and the wider social and ecological systems. The relationship between these systems has been extensively researched and documented in the literature relating to the social acceptance of renewable energies [268,269]. This literature shows that, for example, social acceptance strongly depends on the distance to the turbines and their number [270,271].

### ii. Recommendations for future research

Whilst there is certainly scope to improve the approaches discussed in sections 2 through 4 above, these improvements are mainly constrained within one discipline or area so benefit from an established conceptual framework and common understanding. Arguably improvements in the methods considered within section 5 are more important, because of the relative newness of the field and the need to enhance some basic approaches. But this is also the area in which most disciplines are applied, meaning an additional challenge of a lacking common framework and language. To a large extent, this contradicts the ongoing strive to increase detail and complexity in energy modelling – but increased complexity may also hide the normative and socially constructed nature of the assumptions determining the model output [272].

Based on this discussion, we provide the following suggestions for further work:

- **Resource assessments for onshore wind need to adopt a more self-critical stance,** meaning improving validation of results with measured wind speed/wind turbine data and verifying assumptions with literature, other experts and wider key stakeholders. This means reflecting upon the consequences that the model output and its spatial assumption on wind potentials might trigger, e.g. as stories and narratives for energy policy-making [273,274].
- **Uncertainties surrounding all data inputs** should be explored with sensitivity analyses. Best practice would be to provide a broad range of scenarios which represent the multitude of available options, for example relating to the assumed turbine spacings. Data transparency and public availability of data and models are also key goals for more effective science-policy networking [275].
- In terms of the **technical turbine characteristics**, onshore wind resource assessments should explore the impact of the chosen turbine design(s). Best practice is to employ a state of the art turbine and/or expected future developments in turbine configurations. Where this is not the case, the implications of employing atypical or simplified turbine(s), specific power and/or capacity factors should be quantified.
- In terms of **wind turbine spacing and park planning**, if wind power potentials are assessed over large areas, the extraction of wind energy from the atmosphere may have a significant effect and therefore should be addressed [164]. In addition, the influence of small areas on total potentials should also be checked, using e.g. a sensitivity analysis.
- **Much of the reviewed literature adopts a static viewpoint**, meaning a consideration of one moment in time due to data availability and for other reasons. The dynamic nature of energy transitions in general and wind technologies in particular means that these results quickly become outdated. More useful are studies that develop a modelling framework and/or dataset that can be employed and adjusted by stakeholders for their own analysis (e.g. [9,10]).
- The focus for future research into feasible potentials therefore needs to **explicitly embrace a diversity of approaches and perspectives**, and reflect this in the collaborative research teams. **Holistic frameworks** need to be developed that include multiple dimensions of wind energy impacts as well as their interactions and interdependencies. It is encouraging that two reviewed studies in Table 6 adopt such an MCDA approach, but at the same time an indication of an urgent need for further work.
- **Assessments of wind potentials need to draw up a complete balance sheet**, including all costs and benefits, to be assessed with a systems framework – again, here too there are some promising few examples, but these are limited in number as well as in terms of their focus on specific impact categories.